\documentclass[12pt]{article}

\usepackage{amsmath}
\usepackage{amssymb}
\usepackage{mathtext}
\usepackage{graphicx}

\begin{document}

\title{The restricted two-body problem in constant
curvature spaces.}
\author{Alexey V. Borisov, Ivan S. Mamaev\\
Institute of Computer Science, Udmurt State University\\
1, Universitetskaya str., 426034 Izhevsk, Russia\\
Phone/Fax: +7-3412-500295, E-mail: borisov@rcd.ru, mamaev@rcd.ru
}
\maketitle

\begin{abstract}
We perform the bifurcation analysis of the Kepler problem on $S^3$ and
$L^3$. An analogue of the Delaunay   variables is introduced. We investigate
the motion of a point mass in the field of the Newtonian center moving along
a geodesic  on $S^2$ and $L^2$ (the restricted two-body problem). When
the curvature is small, the pericenter shift is computed using the
perturbation theory. We also present the results of the numerical analysis based on
the analogy with the motion of rigid body.
\end{abstract}


\newpage
\section{Problem statement}

We start with the equations of motion for a particle of unit mass on a
three-dimensional sphere $S^3$ or in a Lobachevsky space $L^3$
(pseudosphere).

The sphere $S^3$ (pseudosphere $L^3$) can be parameterized using the Cartesian
(redundant) coordinates of the four-dimensional
Euclidean space $\mathbb{R}^4$ (the Minkovsky space $\mathbb{M}^4$) with the constraint
\begin{equation}
\label{e1}
\Phi (q)=\frac{1}{2}(g_{\mu \nu}q^\mu q^\nu \pm R^2)=\frac{1}{2}(\langle q,q\rangle\pm R^2)=0
\end{equation}
 where $g=\text{diag} (1,1,1,1)$ ($g=\text{diag}(-1,1,1,1)$) is the
corresponding metrics. Hereinafter, an upper sign in ''$\pm$'' is used for the sphere
and a lower sign is used for the pseudosphere. The  metrics in
$\mathbb{R}^4$ ($\mathbb{M}^4$) generates a metrics in the sphere $S^3$
(the Lobachevsky metrics in the  pseudosphere $L^3$).

In terms of the Cartesian
coordinates, the Lagrangian for the particle's motion  in the field of the potential $U(q)$ is
$$ \mathcal{L} =\frac{1}{2}g_{\mu \nu}\dot{q}^\mu \dot{q}^\nu -U(q),$$
with  the constraint \ref{e1}. Using the Hamiltonian formalism for systems
with constraints (Arnold et al. 1993), we get
\begin{equation}
\label{e2}
 \mathcal{H}=\frac{1}{2}\left( \langle p, p\rangle -\frac{\langle p, q\rangle^2 }
{\langle q, q\rangle }\right) +U(q).
\end{equation}
Then the canonical equations of motion are $\dot q=\displaystyle{\frac{\partial H}{\partial p}} $,
 $\dot p=-\displaystyle{\frac{\partial H}{\partial q}}  $.

 {\bf Two-body problem on $S^3$ ($L^3$).} Consider the two-body problem on the curved
 spaces $S^3$ ($L^3$), where bodies are assumed to be point masses. Let these masses move in
 the field of some potential $U(q_1,q_2)$
 ($q_1, q_2$ are the coordinates of bodies on $S^3$ ($L^3$)).
In this particular case the potential energy $U$ depends on the distance
between two points (this distance is measured along a geodesic). In
our case, there is a center of mass frame of reference such that the two-body problem can be reduced
to the problem of the particle's motion in the field of a fixed attracting
center (i.e. to the Kepler
problem in the case of the Newtonian interaction).  The analogue of the
Kepler problem is superintegrable on $S^3$
($L^3$) (see Kozlov~1994, Borisov~et~al.~1999 and Killing~1885). The
generalization of all the Kepler laws to spaces of constant curvature is
given by Kozlov (1994). But in curved spaces the center-of-mass frame of reference
 does not exist and therefore if the interaction between two bodies is  Newtonian-like,
the two-body problem is not integrable on $S^3$ and $L^3$.

In terms of the Cartesian canonical variables $q_a$ and $p_a$, the
Hamiltonian of the two-body system is (the index $a$ denotes the number of
the mass $m_a$)
$$
 \mathcal{H}=\frac{1}{2m_1}\frac{\langle p_1, p_1\rangle
\langle q_1, q_1\rangle - \langle p_1, q_1\rangle^2}{\langle q_1, q_1\rangle}+$$
\begin{equation}
\label{e3}
\qquad{}+\frac{1}{2m_2}\frac{\langle p_2, p_2\rangle
\langle q_2, q_2\rangle - \langle p_2, q_2\rangle^2}{\langle q_2, q_2\rangle}+U(q_1,q_2).
\end{equation}

{\bf Invariant manifolds.} The three dimensional two-body problem is
rather complicated. Therefore by analogy
with the planar case, we will examine in detail
the motion on the invariant submanifolds of the system . The behavior of the system on
invariant submanifolds allows us to make conclusions
about some properties of the system (nonintegrability, stochasticity)
in the whole phase space. Nevertheless the three-dimensional
problem has not been investigated yet.

The invariant manifolds of the n-body problem in $\mathbb{R}^3$  are planes
and, similarly, if a space is curved, they are spheres $S^2$ (pseudospheres $L^2$).
There is a three-parameter family of such manifolds at any point of $S^3$ ($L^3$)
 (see Borisov et al. 1999).

{\bf Restricted two-body problem.} In the Euclidean space $\mathbb{R}^3$
a passage to the limit is possible in the two-body problem as the mass
of attracting center goes to infinity while the interaction energy remains finite.
There is an inertial frame of reference with origin at the ''heavier'' particle, therefore
if the potential is Newtonian, the restricted problem is the Kepler problem.

Consider a similar passage to the limit on $S^3$ ($L^3$). In this case the attracting
center moves along
a large circle of the sphere (along a geodesic). The second particle (point mass)
moves in the field of attracting center and does not affect the motion of the attracting center.

If the origin is at the first particle, the second  particle moves
in the field of fixed center and gyroscopic forces. The Lagrangian of the system is
\begin{equation}
\label{e4}
\mathcal{L}=\frac{1}{2}\dot{q}^2+\frac{1}{2}\sum\limits_{\mu, \nu}B_{\mu\nu}\dot{q}_{\mu}{q}_{\nu}
+\frac{1}{2}\sum\limits_{\mu, \alpha, \beta}B_{\mu\alpha}{q}_{\alpha}B_{\mu\beta}{q}_{\beta}-V(\mathbf{q}),
\end{equation}
where $\mathbf{B}=||B_{\mu\nu}||$ is the angular velocity matrix of the frame of reference.
Here, $\mathbf{B}\in so(4)$ (i.\,e. it is a skew-symmetric matrix) in the case of $S^3$ and $\mathbf{B}\in so(3,1)$
 in the case of $L^3$.

\section{Bifurcation analysis of the Kepler problem in curved spaces}
Let a particle move in the field of Newtonian-like potential on a sphere $S^3$
(pseudosphere $L^3$).
Using spherical coordinates $q_0=R\cos \theta$, $q_1=R\sin \theta\cos \varphi$,
$q_2=R\sin \theta\sin \varphi\cos \psi$ and $q_2=R\sin \theta\sin \varphi\sin
\psi$,
we can write the Hamiltonian as
\begin{equation}
\label{e5}
\mathcal{H}=\frac{1}{2mR^2}\left(p_{\theta}^2+\frac{1}{\sin^2 \theta}\left(p^2_{\varphi}+
\frac{p_\psi^2}{\sin^2{\varphi}}\right)\right)-\frac{\gamma}{R}\cot\theta.
\end{equation}

Separating variables in (\ref{e5}) gives
\begin{equation}
\label{e6}
\alpha_\psi=p_\psi=\text{const},\quad
\alpha_\varphi^2=p_\varphi^2+\frac{\alpha_\psi^2}{\sin ^2 \varphi}=\text{const},
\end{equation}
\begin{equation}
\label{e7}
E=\frac{1}{2mR^2}\left (p_\theta+\frac{\alpha_\psi^2}{\sin ^2
\theta}\right) -\frac{\gamma}{R}\cot\theta,
\end{equation}
where $\alpha_\psi$ is the projection of the three-dimensional angular
momentum vector $\boldsymbol{M}=m\boldsymbol{q}\times\dot{\boldsymbol{q}}$
(here $\boldsymbol{q}=(q_1,\,q_2,\,q_3)$) onto the axis $q_1$,
$\alpha_\varphi^2$ is the squared momentum $\boldsymbol{M}^2$, $E$ is the energy
constant. It is easy to see that the vector $\boldsymbol{M}$ is an integral of
motion. (In the case of the Lobachevsky plane, all the trigonometric
functions of $\theta$ should be replaced with hyperbolic functions.)

Let us see how the domain of possible motions on $S^3$
($L^3$)
(hereinafter DPM)  depends on the  energy constant $E$ and the  moment constant
$\alpha_\varphi$.

We put $r=R\tan \theta$ ($r=R\tan \theta$) in (\ref{e7}). If $E$ and $\alpha_\varphi$ are fixed
then the DPM are defined as follows
\begin{equation}
\label{e8}
\frac{\alpha_\psi^2}{2m}\left(\frac{1}{r^2}\pm \frac{1}{R^2}\right)
 -\frac{\gamma}{r} \leqslant E .
\end{equation}
Thus to construct a bifurcation diagram we should consider the quadratic equation
\begin{equation}
\label{e9}
\tilde{h}r^2+\gamma r-\frac{\alpha_\varphi^2}{2m}=0,
\end{equation}
where $\tilde{h}=E\pm \displaystyle{\frac{\alpha_\varphi^2}{2mR^2}}$.
The bifurcation set (i.\,e. the locus of $(E,\alpha_\psi)$ at which the domain
of possible motion changes topologically) consists of the curves (see
Fig. 1)
$$
\gamma_1:E=\pm\frac{\alpha^2_\varphi}{2mR^2},\quad
\gamma_2:E=\frac{m\gamma^2}{2\alpha^2_\varphi}\pm\frac{\alpha^2_\varphi}{2mR^2}.
$$
If both roots $r_1$ and $r_2$ of (\ref{e9}) are complex (domain I in Fig. 1)
the motion is impossible. If both roots are real and
positive (domain II), the possible values of $r$ are given by $r_1\leqslant r\leqslant
r_2$. This implies that a particle moves in the ring $\theta_1\leqslant
\theta \leqslant\theta_2$, with $0<\theta_1,\, \theta_2<\displaystyle{\frac{\pi}{2}}$ for  $S^2$.
If the lower root ($r_1$) is negative (domain III), then $r_2\leqslant r$ for the real motion
  on the Lobachevsky plane and $r\leqslant r_1$ for the motion on
a sphere (since $r$ is negative if $\pi/2\leqslant \theta\leqslant \pi$).
It means that on $L^2$ a body moves  exterior to the circle $\theta\leqslant\theta_2$
and if the space is $S^2$, a particle moves in the ring $\theta_1\leqslant \theta\leqslant\theta_2$, where  $0<\theta_1<\pi/2$,
$\pi/2<\theta_2<\pi$.

Note that motions on a sphere are bounded because of
compactness $S^3$. Orbiting time is always
finite. Note also that the ''curved'' Kepler problems are trajectory isomorphic to their
plane analogues, as was shown by Serre (see Appell~1891).

\section{Angle-action variables and analogue of Delaunay variables}
Define the action variables in terms of spherical variables
\begin{equation}
\label{e10}
 I_\psi=\frac{1}{2\pi}\oint p_\psi\, d\psi,\quad I_\varphi=\frac{1}{2\pi}\oint p_\varphi\, d\varphi,
\quad I_\theta=\frac{1}{2\pi}\oint p_\theta\, d\theta,
\end{equation}
where the integral is taken over the whole cycle of the period of motion.

Since $p_\psi=\text{const}$ we have from (\ref{e10}) $I_\psi=p_\psi=\alpha_\psi$.
The kinetic energy in terms of spherical coordinates on $S^3$ is
$T=\displaystyle{\frac{1}{2}}(p_\theta\dot{\theta}+
p_\varphi\dot{\varphi}+p_\psi\dot{\psi})$, and on the invariant sphere $S^2$ where the particle
moves, we have
 $T=\displaystyle{\frac{1}{2}}(p_\theta\dot{\theta}+
\alpha_\varphi\dot{\nu})$; here $\nu$ is a true anomaly (i.e. usual polar
angle).
Equating these two expressions we get $p_\varphi
d\varphi=\alpha_\varphi d\nu - I_\psi d\psi$.
The coordinates $\nu$ and $\psi$ change by $2\pi$ per one revolution of
the orbit. Therefore, after integrating we have
\begin{equation}
\label{e11}
I_\phi=\alpha_\varphi-I_\psi
\end{equation}

To compute the third integral of (\ref{e10}), put $r=R\tan \theta$
($r=R\tan \theta$) and use the equation for the orbit $r(\nu)$ (see
Kozlov 1994; Killing 1885).
\begin{equation}
\label{e12}
r=\frac{p}{1+e\cos \nu},
\end{equation}
where $p=\displaystyle{\frac{\alpha_\varphi^2}{m\gamma}}$ is a parameter of the orbit,
$e=\sqrt{1+\displaystyle{\frac{2\alpha_\varphi^2}{m\gamma^2}\tilde{h}}}$ is the
eccentricity. This implies
\begin{equation}
\label{e13}
I_\theta=\frac{\sqrt{-2m\tilde{h}}}\pi\int^{r_2}_{r_1}\frac{\sqrt{(r-r_1)(r_2-r)}}
{r(1\pm r^2/R^2)}\,dr,
\end{equation}
where $r_1=\displaystyle{\frac{p}{1+e}}$, $r_2=\displaystyle{\frac{p}{1-e}}$.

We get after integration
$$ I_\theta=\sqrt{-2m\tilde{h}}\left(\frac{r_1
\sqrt{r_2^2+R^2}+r_2\sqrt{r_1^2+R^2}}
{\sqrt{2\left(\sqrt{(r_2^2+R^2)(r_1^2+R^2)}+R^2+r_1r_2\right)}}-\sqrt{r_1
r_2}\right)$$ for $S^3$, and

$$ I_\theta=\frac{\sqrt{-2m\tilde{h}}}{2}\left(
\sqrt{(R+r_1)(R+r_2)}-\sqrt{(R-r_1)(R-r_2)}-2\sqrt{r_1
r_2}\right)$$ for $L^3$.

Since $r_1+r_2=-\displaystyle{\frac{\gamma}{\tilde{h}}}$,
$r_1r_2=-\displaystyle{\frac{\alpha_\varphi^2}{2m\tilde{h}}}$, we get with (\ref{e11}) the explicit expression of
the Hamiltonian
\begin{equation}
\label{e18-3.6}
\mathcal{H}=-\frac{m{\gamma}^2}{2(I_{\theta}+I_{\varphi}+I_{\psi})^2}\pm\frac{
(I_{\theta}+I_{\varphi}+I_{\psi})^2}{2mR^2}.
\end{equation}
Similar to the Euclidean space~$\mathbb{R}^3$, the Hamiltonian depends
only on the sum $I_\theta+I_\varphi+I_\psi$, i.e. the
frequencies~$\omega_i=\displaystyle{\frac{\partial H}{\partial I_i}}$,
$i=\theta,\,\varphi,\,\psi$, corresponding to the variables~$I_\theta$,
$I_\varphi$, $I_\psi$, coincide. This is the case of the complete
degeneracy, because all the three-dimensional Liouville~--Arnold tori
foliate into one-dimensional tori i.\,e. circles. Note that unlike the
Hamiltonian in the space~$\mathbb{R}^3$ (see Markeev 1990), expression~\ref{e18-3.6}
has additional terms, which are proportional to~$\displaystyle{\frac1{R^2}}$.

Define new variables~$L$, $G$, $H$, $l$, $g$, $h$ (analogues of the Poincare variables)
\begin{equation}
\label{e18-3.7}
\begin{gathered}
L=I_\theta+I_\varphi+I_\psi,\quad G=I_\varphi+I_\psi,\quad H=I_\psi,\\
l=\omega_\theta,\quad g=\omega_\varphi-\omega_\theta,\quad h=\omega_\psi-\omega_\varphi.
\end{gathered}
\end{equation}

In terms of these variables, the Hamiltonian is
\begin{equation}
\label{e18-3.8}
\mathcal{H}=-\frac{m\gamma^2}{2L^2}\pm\frac{L^2}{2mR^2}.
\end{equation}
With~\ref{e18-3.8} and \ref{e18-3.7}, we have
$$
L=\sqrt{\frac{m\gamma}{-E/\gamma+\sqrt{E^2/\gamma^2\pm 1/R^2}}},\quad
G=\alpha_\varphi,\quad H=\alpha_\psi.
$$

Equation \ref{e18-3.8} implies all the Delaunay   variables except~$l$ are
 integrals of motion. The angle~$l$ is an analogue of the mean
anomaly~$\zeta$ and changes uniformly with the
time~$l=\zeta=\displaystyle{\frac{2\pi}T}(t-\tau)$. Here~$\tau$~is the
time, when the particle passes the pericentre, $T$~is the period of orbit
revolution, which depends only on the energy constant~$E$ (see Killing
1885; Kozlov 1994):
$$
T=\pi\sqrt{\frac m\gamma}R\sqrt{\frac{\pm E/\gamma\pm\sqrt{E^2/\gamma^2\pm 1/R^2}}
{E^2/\gamma^2\pm1/R^2}}.
$$
In terms of the angular length of the orbit's major axis~$a$, the energy
constant~$E=-\displaystyle{\frac\gamma{R\tan a}}$
$\left(E=-\displaystyle{\frac\gamma{R\tan a}}\right)$.

The Delaunay   variables can be expressed in terms of orbit parameters like in
the planar case as it shown by Markeev (1990) and Demin et al. (1999). Choose the angular constants so that if we make gnomonic projection
$g$, $h$
are the images of the pericentre parameter and the longitude of the ascending
node.
Denote them by~$\omega$ and~$\Omega$. Let~$\imath$ be the analogue of orbit
inclination. This value is
equal to the angle between the axis~$q_1$ and the vector~$M$.

Express the variables~$L$, $G$, $H$, $l$, $g$, $h$ in terms of the elements of the orbit~$p$,
$e$, $\imath$, $\tau$, $\omega$, $\Omega$:
\begin{equation}
\label{e18-3.9}
\begin{aligned}
L&=\sqrt{m\gamma R\tan\left(\frac a2\right)},&\quad l&=\zeta,\\
G&=\sqrt{m\gamma p},&\quad g&=\omega,\\
H&=\sqrt{m\gamma p}\cos\imath,&\quad h&=\Omega.
\end {aligned}
\end{equation}
In the case of the Lobachevsky space~$L=\sqrt{m\gamma R\tan\left(\displaystyle{\frac{a}{2}}\right)}$.

\section{Perihelion shift}

The observation of Mercury's perihelion shift is one of the experiments
that proves the general relativity theory (GRT) (see Eddington 1963). This shift arises as a result of
curving of a space near a gravitating body.
Let us prove that in Newtonian mechanics, a Keplerian orbit also
precesses  in a curved space. Although the laws of precession in these theories are different.
We will take the restricted two-body problem as a model problem. This
problem is not integrable but if the velocity of the heavier particle is low
we can analyze the problem using the perturbation theory. Here we do not mean to give
a new physical justification of the perihelion shift, already given in GRT
and accepted as classical. We just point out that some phenomena of the practical Celestial
Mechanics admit another interpretations (together with the planet
nonsphericity,
atmosphere refraction and so on). The addition of curvature to the classical Newtonian
mechanics is an example of such interpretations.

Consider the restricted two-body problem on~$S^2$ $(L^2)$. As usual, the
sphere (pseudosphere) is assumed to be embedded in~$\mathbb{R}^3$
$(\mathbb{M}^3)$:
$\{\boldsymbol{q}=(x,\,y,\,z)|\langle\boldsymbol{q},\,\boldsymbol{q}\rangle=x^2+y^2\pm
z^2=\pm R^2\}$. Let an attracting center move along the geodesic on
$xz$~plane, and we choose the (moving) frame of reference that the
attracting center is at the  north pole of the sphere
(pseudosphere)~$\boldsymbol{e}_3=(0,\,0,\,1)$. The Lagrangian of point
mass (particle of unit mass) is
\begin{equation}
\label{e18-4.1}
\mathcal{L}=\frac{1}{2}\langle\boldsymbol{\dot
q},\,\boldsymbol{\dot q}\rangle+\gamma\frac{\langle\boldsymbol{e_3},\,\boldsymbol{q}\rangle}
{\sqrt{R^2\mp\langle\boldsymbol {e}_3,\,\boldsymbol {q}\rangle^2}}+\langle\boldsymbol {\dot q},\,{\bf
B}\boldsymbol {q}\rangle+ \frac{1}{2}\langle{\bf B}\boldsymbol {q},\,{\bf B}\boldsymbol {q}\rangle,
\end{equation}
here~$\bf
B$~is the angular velocity matrix of the frame of reference,
$$
{\bf B}=\left(\begin{matrix}
0 & 0 & w \\
0 & 0 & 0 \\
\mp w & 0 & 0
\end{matrix}\right).
$$

Let us assume that the typical size of the domain of motion of the point mass is small
in comparison with the radius of curvature~$R$. Then we can analyze the problem using perturbation theory.
 Suppose also that the angular
velocity of the attracting center's motion is small in comparison with the rotation frequency
of the point mass moving along the corresponding Keplerian orbit.
Take the length $r=R\tan
\theta$ $(r=R\tan\theta)$ and azimuth angle~$\varphi$ as the coordinates on a sphere (pseudosphere)
and transform \ref{e18-4.1} as
\begin{equation}
\label{e18-4.2}
\mathcal{L} = \frac{1}{2}\Biggl(\frac{\dot
r^2}{\Bigl(1{\pm}\displaystyle{\frac{r^2}{R^2}}\Bigr)^2}  + \frac{{r^2}{{\dot {\varphi}}^2}}{1{\pm}\displaystyle{\frac{r^2}{R^2}}}\Biggr)
+\frac{\gamma}{r} +
2\frac {w}{R}\frac{r^2\dot r}{\Bigl(1{\pm}\displaystyle{\frac{r^2}{R^2}}\Bigr)^2}\cos\varphi{\mp}
\frac{w^2}{2}\frac{r^2}{1{\pm}\displaystyle{\frac{r^2}{R^2}}}\sin^2\!\varphi.
\end{equation}
Here~$w=\displaystyle{\frac {v}{R}}$, and $v$~is the linear velocity of
the motion of the noninertial frame of reference. If~$\displaystyle{R\rightarrow\infty}$,
the problem is reduced to the planar Kepler problem.

The terms in \ref{e18-4.2}, which are linear with respect to the
velocity, have the order~$\displaystyle{\frac1{R^2}}$ and can not be omitted. To study
the evolution of the orbit's shape in the unperturbed Kepler problem we
express the equations \ref{e18-4.2} in terms of~$p$, $\omega$, $e$,
$\varphi$. Here,~$e$~is the eccentricity, $\omega$~is the longitude of the
orbit's pericenter, $\varphi$~is the azimuth angle, $p$~is the orbit's
parameter, associated with the energy~$E$ of the unperturbed Kepler
problem by following
\begin{equation}
\label{e18-4.3}
E=-\frac{1-e^2}{2p}\pm\frac p{2R^2}.
\end{equation}
The new variables are expressed in terms of coordinates and velocities
(hereinafter~$\gamma=1$)
\begin{equation}
\label{e18-4.4}
r=\frac p{1+e\cos(\varphi-\omega)},\quad\frac{r^2}{1\pm\displaystyle{\frac{r^2}{R^2}}}\dot r=
\frac{e\sin(\varphi-\omega)}{\sqrt
p},\quad\frac{r^2\dot\varphi}{1\pm\displaystyle{\frac{r^2}{R^2}}}=\sqrt p.
\end{equation}
Hereinafter, for the sake of simplicity,  we don't substitute the expression for~$r$
in terms of~$p$, $e$, $\omega$, $\varphi$. The Poisson brackets for~$p$, $e$, $\omega$,
$\varphi$ are
\begin{equation}
\label{e18-4.5}
\begin{aligned}
\{p,\,e\}&=-\frac{4w}R\frac{pr^2}{1\pm\displaystyle{\frac{r^2}{R^2}}}\sin\varphi\sin(\varphi-\omega);\\
\{p,\,\omega\}&=-2\sqrt p+\frac{4w}R\frac{pr^2}{1\pm\displaystyle{\frac{r^2}{R^2}}}\sin\varphi\cos
(\varphi-\omega);\\
\{e,\,\omega\}&=\frac{1-e^2+\displaystyle{\frac{p^2}{R^2}}}{e\sqrt p}+\frac{4w}R\frac{pr}
{e\left(1\pm\displaystyle{\frac{r^2}{R^2}}\right)}\sin\varphi;\\
\{p,\,\varphi\}&=-2\sqrt p;\\
\{e,\,\varphi\}&=-\frac{2\cos(\varphi-\omega)+e+e\cos^2(\varphi-\omega)}{\sqrt p};\\
\{\omega,\,\varphi\}&=-\frac{\sin(\varphi-\omega)(2+e\cos(\varphi-\omega))}{e\sqrt p},
\end{aligned}
\end{equation}
and the Hamiltonian is
\begin{equation}
\label{e18-4.6}
\mathcal{H} = -\frac{1-e^2}{2p}\pm\frac p{2R^2}\pm\frac{w^2}2\frac{r^2\sin^2\varphi}
{1\pm\displaystyle{\frac{r^2}{R^2}}}.
\end{equation}
Expressions  \ref{e18-4.5} and \ref{e18-4.6} imply that
when~$R\rightarrow\infty$, the variables~$p$, $e$, $\omega$~are slow,
$\varphi$~is fast. To define the secular change of the orbit's parameters,
when~$R\gg r$, we neglect the terms with order higher than~$\displaystyle{\frac1{R^2}}$
and average the equations of motion over the period of unperturbed motion.
Averaging over the period is equal to averaging over~$\varphi$ with a
weight
function~$\displaystyle{\frac1{2\pi}}\int_0^{2\pi}f(\varphi)\rho(\varphi)d\varphi$. Here
the weight function $\rho$ is defined by the derivative~$\dot{\varphi}$
from~\ref{e18-4.4} as~$\rho=\displaystyle{\frac1{\dot\varphi}}$.

So, we have the system:
\begin{equation}
\label{e18-4.7}
\begin{aligned}
\dot p&=\mp\frac v{R^2}\frac{2ep^{7/2}}{(1-e^2)^{5/2}}
\left(\cos\omega+\frac52 v\frac{e\sqrt p\sin\omega\cos\omega}{1-e^2}\right),\\
\dot e&=\pm\frac v{R^2}\frac{p^{5/2}}{(1-e^2)^{3/2}}\left(\cos\omega+\frac52v
\frac{e\sqrt p\sin\omega\cos\omega}{1-e^2}\right),\\
\dot\omega&=\mp\frac v{R^2}\frac{p^{5/2}}{(1-e^2)^{5/2}}\left(\frac{1-2e^2}e
\sin\omega+v\sqrt p\left(2-\frac52\cos^2\omega\right)\right).
\end{aligned}
\end{equation}

The equations \ref{e18-4.7} has the integral
\begin{equation}
\label{stars0}
-\frac{1-e^2}{2p}=C,
\end{equation}
This integral implies that there is no secular change of the  energy of
the unperturbed system \ref{e18-4.3} to this approximation (Laplace's theorem).

The phase portrait of \ref{e18-4.7} on the surface of the integral
\ref{stars0} depends on the parameter~$b=\displaystyle{\frac v{\sqrt C}}$. Figure 2 shows the projection
of the trajectories onto~$(\omega,\,e)$ plane for different values of~$b$.
The parameter $b$ describes the ratio of the velocity of the attracting
center to the characteristic  velocity of the particle along the Keplerian orbit.
The equation \ref{e18-4.7} implies that the curvature sign determines the direction of
the motion along the trajectory but not the shape of the trajectory.
The value of curvature defines the velocity of motion along the trajectory.

It is clear from the figures
that the velocity of perihelion shift depends not only on the eccentricity of the orbit
but also on the orientation of the orbit with respect to the direction of motion of the attracting center.

If~$b$ is small, there exist two stable periodic orbits with the non-zero
eccentricity. The main axis of the orbits is perpendicular to the
direction of the attracting center's motion. At pericenter, the direction
of the point mass motion along one of the orbit coincides with the
direction of the attracting center's
motion~$\left(\omega=\displaystyle{\frac\pi2}\right)$. When the particle
moves along another orbit ~$\left(\omega=\displaystyle{\frac{3\pi}2}\right)$ at the
pericenter, its direction is opposite to the direction of the attracting
center's motion. When~$b$ increases the orbit with~$\omega=\displaystyle{\frac{3\pi}2}$
becomes unstable, and if~$b$ is sufficiently large, the stable orbit
with~$\omega=\displaystyle{\frac{\pi}2}$ disappears.\looseness=1

The projections of the trajectories of the non-averaged system onto~$(\omega,\,e)$ plane are also shown in the
figures. Here, we can see small oscillations (for the variables~$\omega$, $e$) near the trajectories
of the averaged system (see~Fig.~3).

Remind that standard explanation of the perihelion shift is based on the
Schwarzschield solution (Eddington 1963) and implies that the shift velocity does
not depend on the orientation of the orbit $\omega$. This is not the case
in our problem statement. There always exist orbits such that only their
eccentricities change their values but their pericenters have almost no
shift. Moreover, there are fixed points of the system (\ref{e18-4.7}),
corresponding to the periodic orbits which do not change their form and
orientation. Note that the Schwarzschield-like metrics can be constructed
if  boundary conditions (at infinity) correspond to the space of constant
curvature as it shown by Chernikov (1992). And also the restricted two-body problem can be
generalized for such metrics. It is clear that the velocity of the
perihelion shift depends on both~$\omega$ and $e$.

\section{Isomorphism with the spherical top dynamics}

Consider the restricted two-body problem on the sphere~$S^2$ in the general case without assumption that
the curvature is small.
We define the new variables~$\boldsymbol{M}$, $\boldsymbol{\gamma}$ using the map~${T^*R^3\rightarrow
e(3)}$,
by
\begin{equation}
\label{e18-5.1} \boldsymbol{\gamma}=\frac{\boldsymbol
{q}}R,\quad\boldsymbol{M}=\boldsymbol{\gamma}\times\boldsymbol{p}.
\end{equation}
Here, the canonical Poisson brackets~$\{q_i,\,p_j\}=\delta_{ij}$ are transformed to the Lie~--Poisson
brackets
corresponding to~$e(3)$ algebra. The equations of motion are
\begin{equation}
\label{e18-5.1_12}
\boldsymbol{\dot M}=\boldsymbol{M}\times\frac{\partial\mathcal{H}}{\partial\boldsymbol{M}}+
\boldsymbol{\gamma}\times\frac{\partial\mathcal{H}}{\partial\boldsymbol{\gamma}},\quad
\dot{\boldsymbol{\gamma}}=\boldsymbol{\gamma}\times\frac{\partial\mathcal{H}}{\partial\boldsymbol{M}},
\end{equation}
where the Hamiltonian is
\begin{equation}
\label{e18-5.1_34}
\mathcal{H}=\frac{1}{2}\boldsymbol{M}^2+(\boldsymbol{M},\,\boldsymbol{w})+U(\boldsymbol{\gamma}),\quad
U(\boldsymbol{\gamma})=-\frac{\gamma_3} {\sqrt{\gamma^2_1+\gamma^2_2}}.
\end{equation}
These equations (see Borisov et al. 2001) have two integrals of motion: the area
integral~$(\boldsymbol{M},\,\boldsymbol{\gamma})=C$ and the geometric
integral~$(\boldsymbol{\gamma},\,\boldsymbol{\gamma})=1$. In our case~$C=0$. Note that this
system describes the motion of spherical top in the
potential~$U(\boldsymbol{\gamma})$ and in the field of gyroscopic
forces.\looseness=-1

\enlargethispage*{\baselineskip}

For~$\boldsymbol{w}=0$ this system is (super)integrable (the Kepler
problem on~$S^2$), and in this case simple geometric interpretation of the
motion exists: {\it the variable~$M_3=\text{const}$ and the projection of the
trajectory onto the plane~$(M_1,\,M_2)$ is a circle shifted from the
origin} (incidentally, Hamilton noticed a similar thing, in the
two-dimensional Kepler problem). Indeed in the consequence of
$(\boldsymbol{M},\vec{\gamma})=0$ we have
$\boldsymbol{M}=\dot{\vec{\gamma}}\times\gamma$ and using the equation
(\ref{e12}) we obtain
$$\gamma_1=\frac{p\cos \nu}{\sqrt{p^2+(1+e\cos \nu)^2}},
\quad \gamma_2=\frac{p\sin \nu}{\sqrt{p^2+(1+e\cos \nu)^2}},$$
$$ \gamma_3=\frac{1+e\cos \nu}{\sqrt{p^2+(1+e\cos \nu)^2}},
\quad \frac{p^2e\dot{\nu}\sin \nu}{(p^2+(1+e\cos \nu)^2)^{3/2}}=M_3=const $$
where $\nu$ is a longitude on the sphere $\vec{\gamma}^2=1$. Eliminating $\dot{\nu}$ from the last equation
we can get
$$M_1=-p^{-1}M_3(e+\cos \nu),\quad M_2=-p^{-1}M_3\sin \nu $$

For~$\boldsymbol{w}\ne0$, according to the Liouville--Arnold theorem,  an
additional integral must exist the system to be completely integrable. We
will show soon that in the general case, the additional integral does not
exist (see also Borisov et al. 1999, Cherno\"{\i}van et al. 1999).

We construct the Poincare map to  study the problem numerically.
To construct it we use the analogy with the motion of rigid body and
choose the Andoyer canonical variables~$(L,\,G,\,l,\,g)$ according to Borisov et al. (2001) as
$$
\begin{gathered}
L=M_3,\quad l=\arctan\Bigl(\frac{M_2}{M_1}\Bigr),\\[-1mm]
G=(\boldsymbol{M},\boldsymbol{M}),\quad g=\arccos\left(\frac{-\gamma_3}
{\sqrt{1-M^2_3/(\boldsymbol{M},\boldsymbol{M})}}\right).
\end{gathered}
$$
\goodbreak
\noindent
Then  the Hamiltonian as a function of these coordinates is
\vskip-0.5mm\noindent
$$
\mathcal{H}=\frac12G^2+w\sqrt{G^2-L^2}\cos l+\frac{\sqrt{G^2-L^2}\cos
g}{\sqrt{G^2\sin^2 g+L^2\cos^2 g}}.
$$
\vskip-1.5mm\noindent
The equations of motion are canonical:
\vskip0.5mm\noindent
\begin{equation}
\label{e18-5.2}
\dot l=\frac{\partial\mathcal{H}}{\partial L},\quad\dot L=-\frac{\partial\mathcal{H}}{\partial l},\quad
\dot g=\frac{\partial\mathcal{H}}{\partial G},\quad\dot G=-\frac{\partial\mathcal{H}}{\partial g}.
\end{equation}
\vskip-0.5mm\noindent

Let us fix the energy level $\mathcal{H}=E$ and define the Poincare
section by the relation~$g=\displaystyle{\frac\pi2}$. On this
two-dimensional surface we choose the variables~$\displaystyle{\frac LG}$ and~$l$ as the
coordinates of the Poincare map (similarly to Borisov et al. 2001). The domain of
definition of the variables is compact: $l\bmod2\pi$, $\displaystyle{\left|\frac
LG\right|\leqslant 1}$ and the flow (\ref{e18-5.2}) defines corresponding Poincare
map.

By direct substitution into \ref{e18-5.2}, it is easily proved that
$$
\begin{aligned}
\dot L(-L,\,-l,\,G,\,g)&=-\dot L(L,\,l,\,G,\,g),&\quad
\dot l(-L,\,-l,\,G,\,g)&=-\dot l(L,\,l,\,G,\,g),\\
\dot G(-L,\,-l,\,G,\,g)&=\dot G(L,\,l,\,G,\,g),&\quad
\dot g(-L,\,-l,\,G,\,g)&=\dot g(L,\,l,\,G,\,g).
\end{aligned}
$$
So, each trajectory $C_1$ with the initial conditions~$(L_0,l_0,$ $G_0,g_0)$
corresponds to a similar trajectory~$C_2$ with initial conditions~$(-L_0,\,-l_0,\,G_0,\,g_0)$,
and each point~$(L,\,l,\,G,\,g)$ of~$C_1$ corresponds to
a point~$(-L,\,-l,\,G,\,g)$ of~$C_2$. This means the
Poincare map (for the chosen section $g=\displaystyle{\frac\pi2}$ is central symmetric. The phase portraits for the different
values of the energy~$E$ and parameter~$w$ are shown in Fig.~4.

\enlargethispage*{\baselineskip}

It is easy to see in the figures that the stochastic layer increases
as
the  energy~$E$. This proves that the two-body problem in general case is not integrable.
 The fixed points in Fig.~4 correspond to periodic
trajectories of the particle, which play an impotent role in the qualitative analysis of the system.

After the publication of the book by Borisov et al. (1999) and paper by
Cherno\"{\i}van et al. (1999),  at our suggestion, S.\,L.\,Ziglin could
prove that the additional meromorphic integral does not exist for the
potentials that are the analogous to the Newtonian and Hooke interaction
for any value of the parameters (see Ziglin 2001 and Ziglin 2003). \vspace*{-3mm}

\section{Hill domains and relative equilibrium}
With the equation of motion \ref{e18-5.1_12}, the integral of energy of
the restricted problem \ref{e18-5.1_34} can be written as
$$
\begin{gathered}
\frac{1}{2R^2}\boldsymbol{\dot
q}^2+U_{*}(\theta,\varphi)=\mathcal{E}=\text{const},\\
U_{*}=\frac{1}{2}\sin^2\theta\sin^2\varphi-\mu\cot\theta,\quad\mu=\frac\gamma{w^2}>0,\quad
\mathcal{E}=\frac{E}{w^2}+\frac12,
\end{gathered}
$$
where $\theta$, $\varphi$ are the spherical
coordinates on $S^2$ (it means that $\theta$ is a latitude and $\varphi$
is a longitude).
\goodbreak

If the energy $E$ is fixed (therefore,~$\mathcal{E}$ is also constant)
the domain of motion on $S^2$ is defined by
\begin{equation}
\label{stars1}
\mathcal{E}-U_{*}(\theta,\varphi)\geqslant 0.
\end{equation}
By analogy with the classical restricted three-body problem (Arnold et
al. 1993), we will call this domain {\it the Hill domain of the restricted
two-body problem on a sphere}.

When the parameters of the system are fixed, the shape of the Hill domains
is defined by the singularities and critical points of the effective
potential~$U_{*}(\theta,\varphi)$. The singularities of $U_{*}$ are at the
poles of a sphere. And, since $\mu>0$ and
$-\mu\cot\theta\xrightarrow[\theta\to 0]{}\infty$,
Hill domain is always not empty, because near $\theta=0$ the inequality
(\ref{stars1}) holds. Each critical point of the
function~$U_{*}(\theta,\varphi)$ corresponds to the equilibrium position
of the particle (here, the frame of reference rotates with the attracting
center). This equilibrium position is usually called {\it a relative
equilibrium}. Note that (we will prove this below) in the fixed frame of
reference the attracting center moves along the large circle and the
particle moves along the another circle parallel to the large circle, with
it being at the same meridian with the attracting center.

Find the location of critical points by solving the system
\begin{equation}
\label{stars4}
\frac{\partial
U_{*}}{\partial\theta}=\sin\theta\cos\theta\sin^2\varphi+\frac\mu{\sin^2\theta}=0,\quad
\frac{\partial U_{*}}{\partial \varphi}=\sin^2\theta\sin\varphi\cos\varphi.
\end{equation}
We obtain the following results:

\noindent 1$^\circ$ {\it if
$0<\mu<\mu_{*}=\displaystyle{\frac{3\sqrt3}{16}}$, there are four critical
points on the meridians $\varphi=\displaystyle{\frac\pi2}$ and
$\varphi=\displaystyle{\frac32\pi}$ (two points on each meridian). Their
latitudes~$\displaystyle{\frac{\pi}{2}}<\theta_1<\theta_2<\pi$ are defined by the equation}
$$
\frac12\sin2\theta+\frac\mu{\sin^2\theta}=0;
$$

\noindent 2$^\circ$
{\it if $\mu>\mu_{*}$, the function $U_{*}(\theta,\varphi)$ has no critical
points at all}.

\noindent It is easy to see, that in the case $1^\circ$ the critical
points $\left(\displaystyle{\frac\pi2,\theta_2}\right)$,
$\left(\displaystyle{\frac{3\pi}2,\theta_2}\right)$ are the saddle points
of the function $U_{*}$, and the points $\left(\displaystyle{\frac\pi2,\theta_1}\right)$,
$\left(\displaystyle{\frac{3\pi}2,\theta_1}\right)$ are the strict maxima.

Hill domains for both cases (1$^\circ$ and
2$^\circ$) are shown  in Fig. \ref{X1}, \ref{X2}.
It is clear from Fig. \ref{X1} that fixed points are in the semisphere
opposite to the attracting center. We will use linear approximation to investigate the stability of the obtained fixed
points (relative equilibria).
Let the point~$(\varphi_i,\theta_j)$, $i,j=1,2$, be the corresponding fixed point, where
$\varphi_i\in\{\displaystyle{\frac\pi2,\frac32\pi\}}$,
$\theta_j\in \{\theta_1,\theta_2\}$. According to \ref{stars4}, it is very convenient
to parameterize $\mu$ by the latitude of the fixed point:
\begin{equation}
\label{stars2}
\mu=-\cos\theta_j\sin^3\theta_j.
\end{equation}
Since $\mu>0$ for attracting center,
$\theta_j\in\left[\displaystyle{\frac\pi2},0\right)$, moreover for $j=1$
$\displaystyle{\frac\pi2}<\theta_1<\theta_{*}$ and these points correspond to the maximum of
the effective potential~$U_{*}$ and $\theta_{*}<\theta_2<\pi$ to the saddle
point. Here, $\theta_{*}$ denotes the value of $\theta$, for which $\mu$
\ref{stars2} reaches the maximum $\mu=\mu_{*}=\displaystyle{\frac{3\sqrt3}{16}}$.

Let us introduce canonical impulses $p_\theta$, $p_\varphi$ corresponding
to the spherical angles. In the fixed points their values are
$p_\theta=0$, $p_\varphi=\pm w\cos\theta_j\sin\theta_j$. Expand the Hamiltonian \ref{e18-5.1_34}
in the vicinity of the fixed point up to the second power, using the following
canonical variables:
$$
p_\theta=X,\quad p_\varphi=\pm w\cos\theta_j\sin\theta_j+Y,\quad\varphi=\varphi_i+y,\quad\theta=\theta_j+x.
$$
We obtain
$$
\begin{aligned}
H=H_0+\frac12\left(X^2+\frac{Y^2}{\sin^2\theta_j}\right)+
w\left(yX-\frac{\cos^2\theta_j}{\sin^2\theta_j}xY\right)+\\
+\frac12w^2\cos^2\theta_j\left(\frac{x^2}{\sin^2\theta_j}+y^2\right)+\dots,\quad
H_0=\text{const},\quad j=1,2.
\end{aligned}
$$
The eigenvalues of the corresponding linearized system are
\begin{equation}
\label{stars3}
\begin{gathered}
\lambda_{1,2}=\pm
w\sqrt{\frac{1-\cos\theta_j-2\cos^2\theta_j}{1-\cos\theta_j}},\quad
\lambda_{3,4}=\pm w\sqrt{\frac{1+\cos\theta_j-2\cos^2\theta_j}{1+\cos\theta_j}},\\ j=1,\,2.
\end{gathered}
\end{equation}

The study of the radical expressions in \ref{stars3} gives us the following results: for
$\theta=\theta_1$ and $\theta=\theta_2$, $\lambda_{3,4}$ are always real and  $\lambda_{1,2}$
are real for $\theta=\theta_1$ and purely imaginary for $\theta=\theta_2$.

This means that

{\it relative equilibria in the restricted two-body problem on a sphere are always unstable}.

Note that, according to the theorem of central manifold, the existence of two purely imaginary eigenvalues for the points
$\left(\displaystyle{\frac\pi2},\theta_2\right)$ and $\left(\displaystyle{\frac32\pi},\theta_2\right)$
results  in the existence of an unstable (hyperbolic)
periodic solution near these points. Fig. \ref{X3} shows these solutions for different values of energy.

\section{Acknowledgements}
This work was supported by RFBR (04-05-64367 and
05-01-01058), CRDF (RU-M1-2583-MO-04),
INTAS (04-80-7297) and the program ``State Support for Leading
Scientific Schools'' (136.2003.1).

\newpage
\begin{figure}[!ht]
\begin{center}
\includegraphics{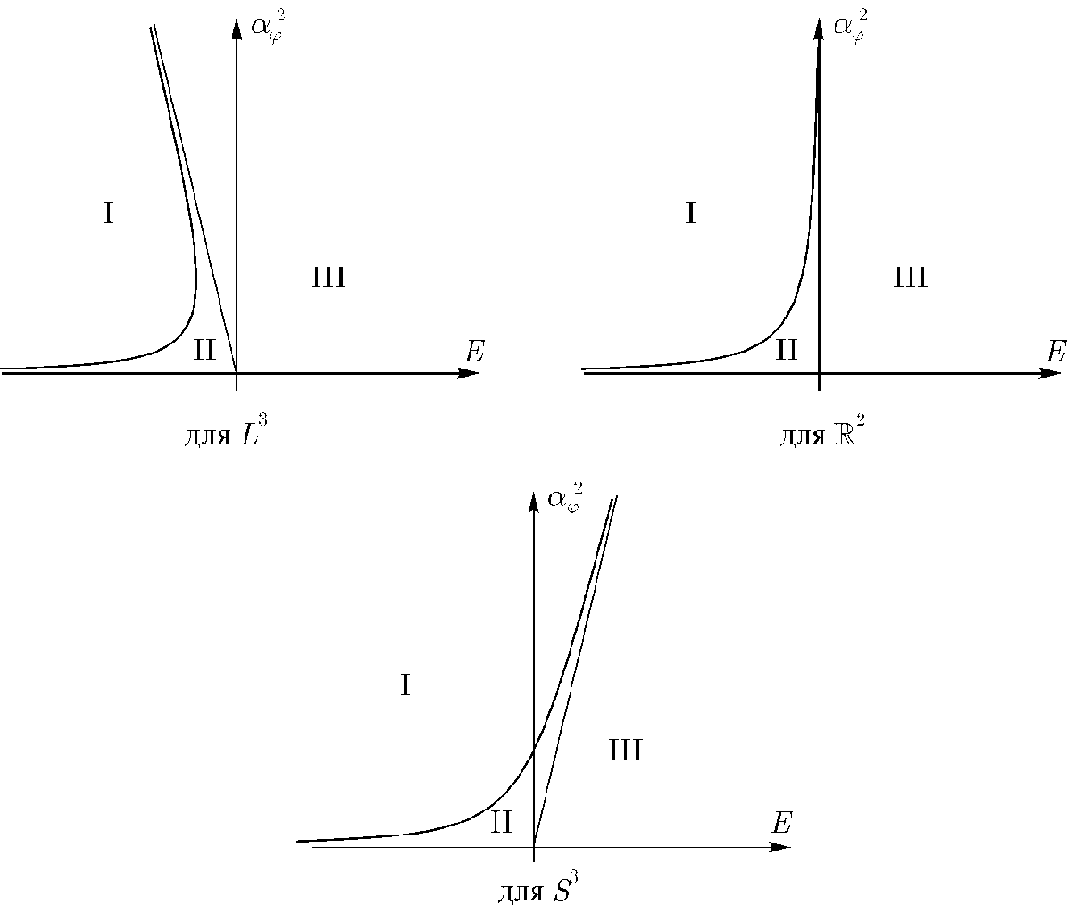}
\end{center}
\caption{The bifurcation diagrams of the Kepler problem}
\end{figure}

\newpage
\begin{figure}[!ht]
\begin{center}
\begin{tabular} {cc}
\includegraphics{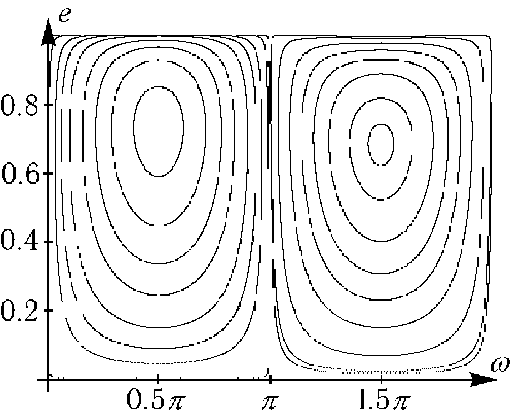}$b=0.1$ &
\includegraphics{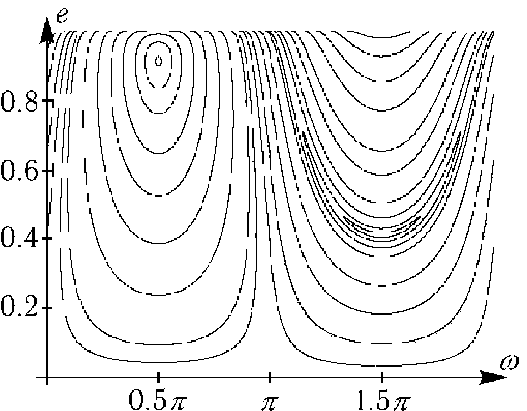} $b=1.26$\\
\includegraphics{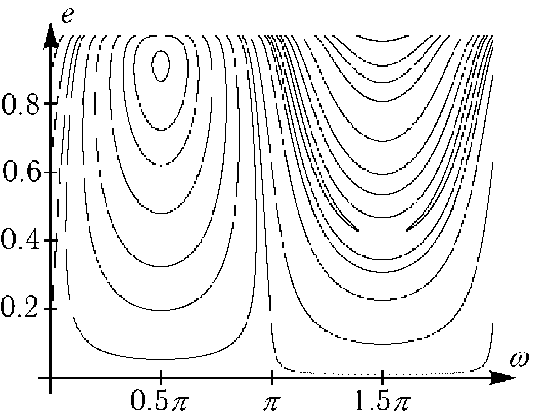} $b=1.27$&
\includegraphics{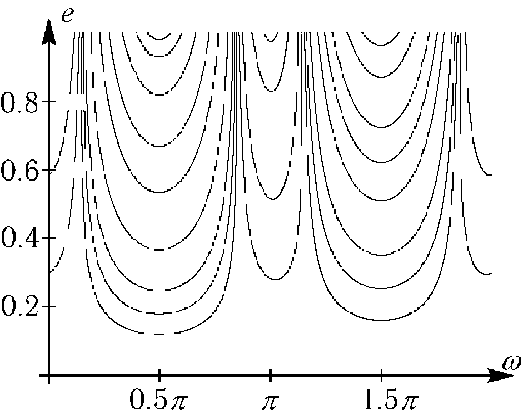}$b=25$\\
\end{tabular}
\end{center}
\vspace*{-2mm}
\caption{Phase portraits of the averaged system}
\end{figure}

\newpage
\begin{figure}[!hp]
\begin{center}
\begin{tabular}{cc}
\includegraphics{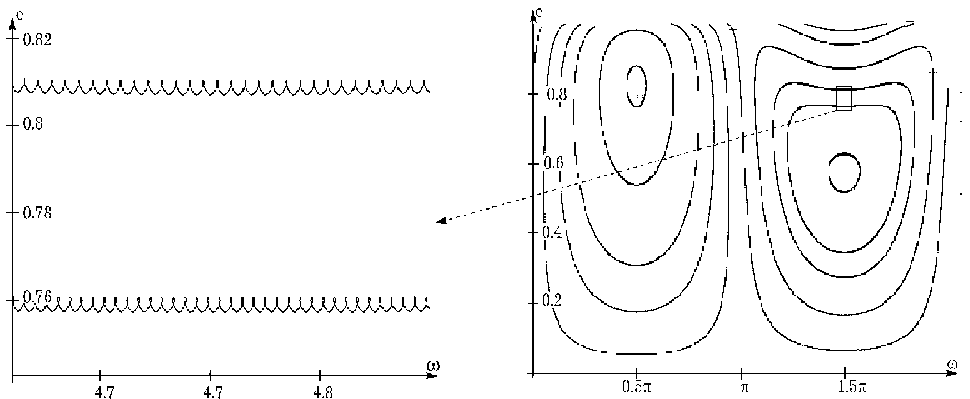} $b=0.5$\\
\includegraphics{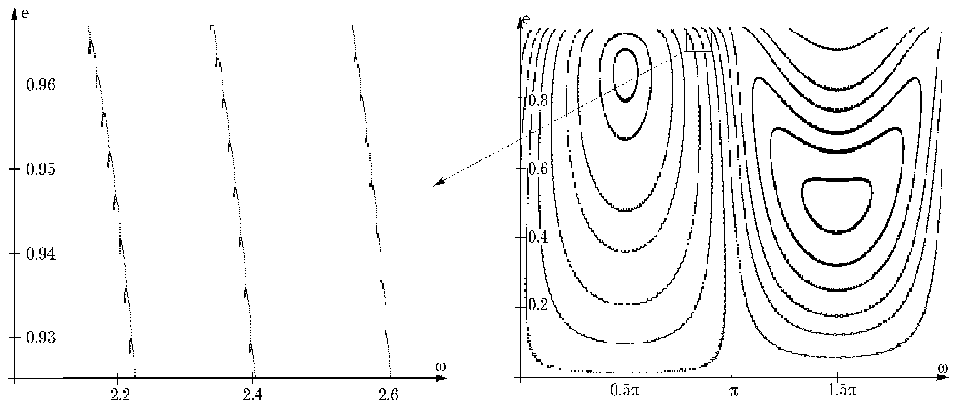} $b=0.9$\\
\includegraphics{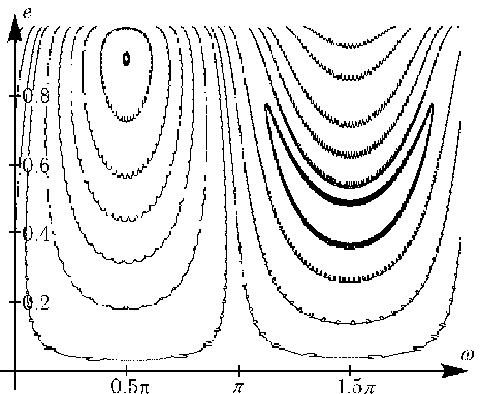} $b=1.2$\\
\includegraphics{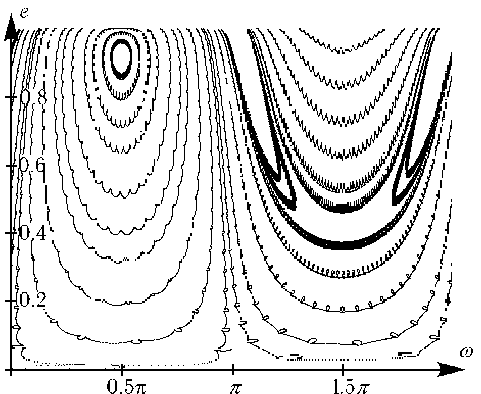} $b=1.5$\\
\end{tabular}
\end{center}
\caption{Phase portrait of the non-averaged system}
\end{figure}

\begin{figure}[!hp]
\centering
\begin{tabular}{c}
\includegraphics{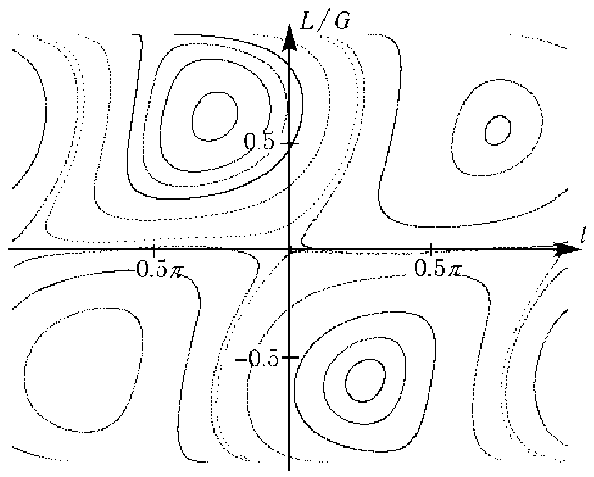} $E=1$, $w=0.1$\\
\includegraphics{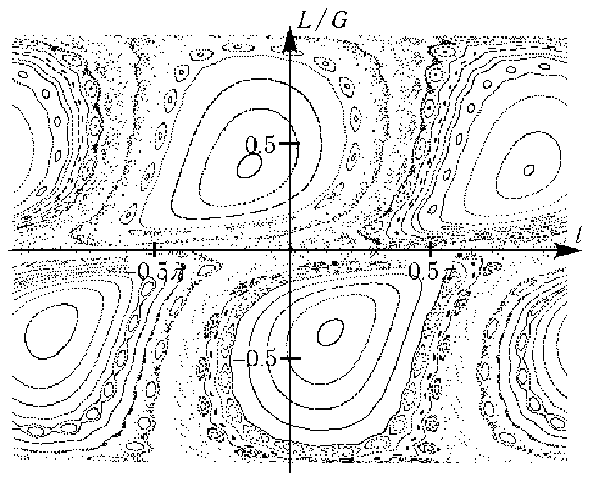} $E=3$, $w=0.1$ \\
\includegraphics{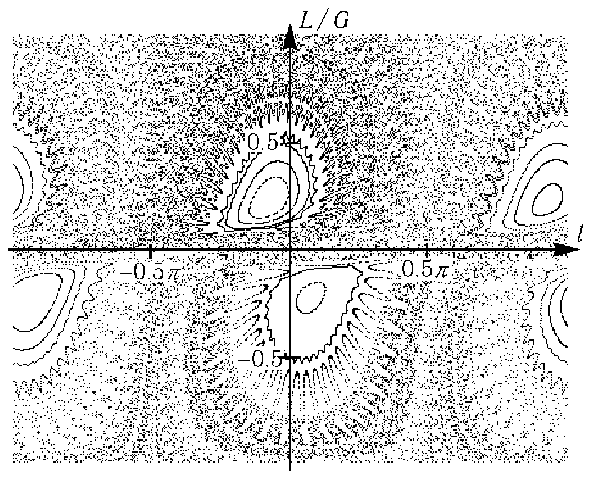} $E=10$, $w=0.1$ \\
\end{tabular}
\caption{The Poincare map for the various energy values and for $w\ne0$}
\label{fig4}
\end{figure}

\begin{figure}[!ht]
\begin{center}
\begin{tabular}{cc}
\includegraphics{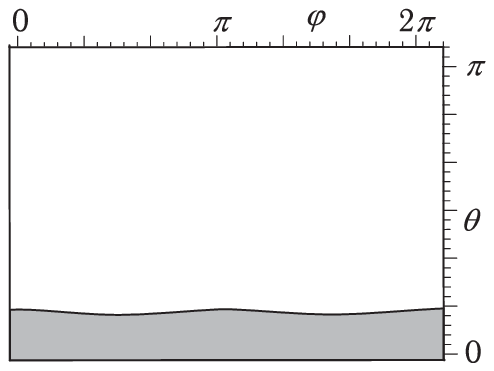} $\mathcal{E}=-0.5$ & \includegraphics{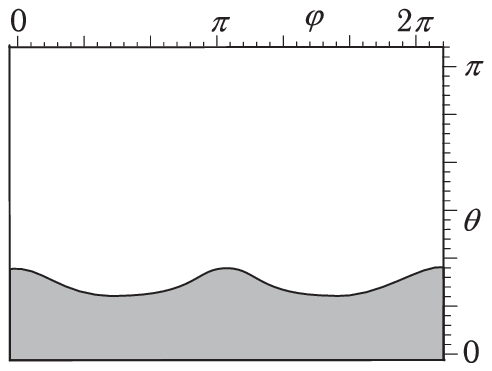} $\mathcal{E}=-0.2$\\
\includegraphics{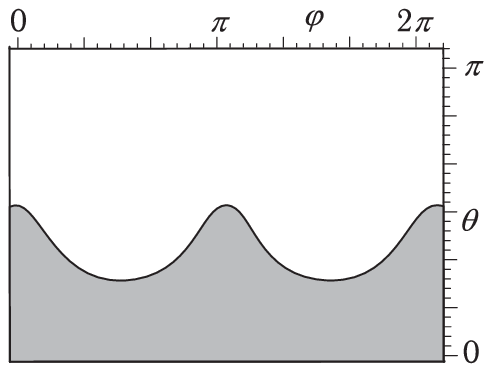} $\mathcal{E}=0$ &  \includegraphics{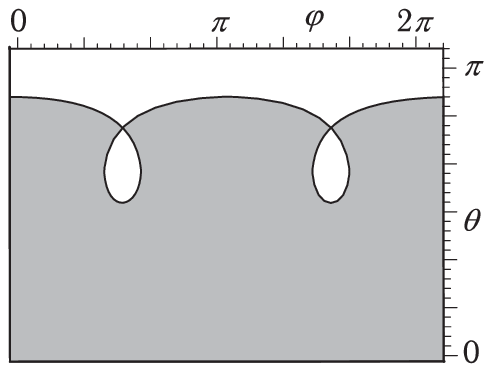} $\mathcal{E}=0.5$\\
\includegraphics{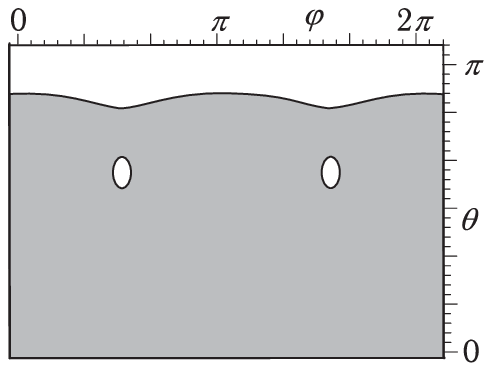} $\mathcal{E}=0.525$ & \includegraphics{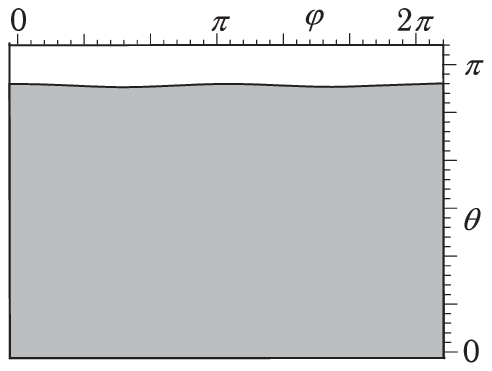} $\mathcal{E}=0.7$ \\
\end{tabular}
\end{center}
\caption{In the case 1$^\circ$ Hill domains (gray shade)  for
$\mu=0.25<\mu_{*}$ and $w=2.0$}
\label{X1}
\end{figure}

\begin{figure}[!ht]
\begin{center}
\begin{tabular}{cc}
\includegraphics{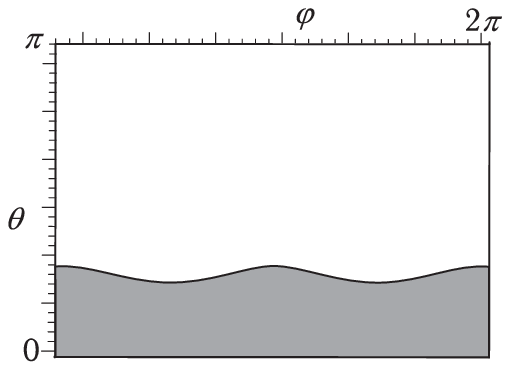} $\mathcal{E}=-0.5$  \\
\includegraphics{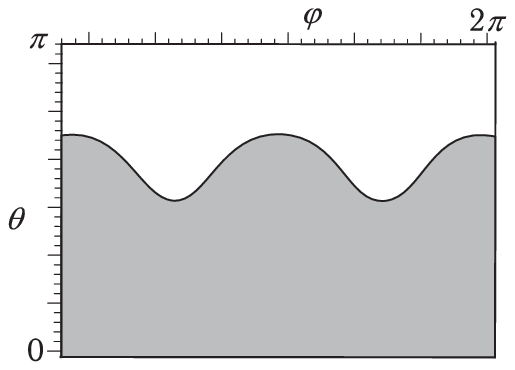} $\mathcal{E}=0.5$ \\
\includegraphics{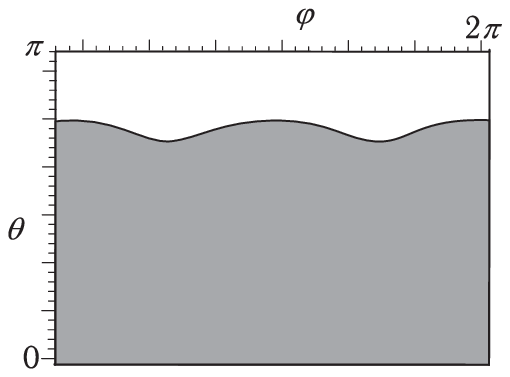} $\mathcal{E}=0.8$ \\
\end{tabular}
\end{center}
\caption{In the case 2$^\circ$ Hill domains (gray shade)  for $\mu=0.6>\mu_{*}$ and $w=2.0$}
\label{X2}
\end{figure}

\begin{figure}[!ht]
\begin{center}
\includegraphics{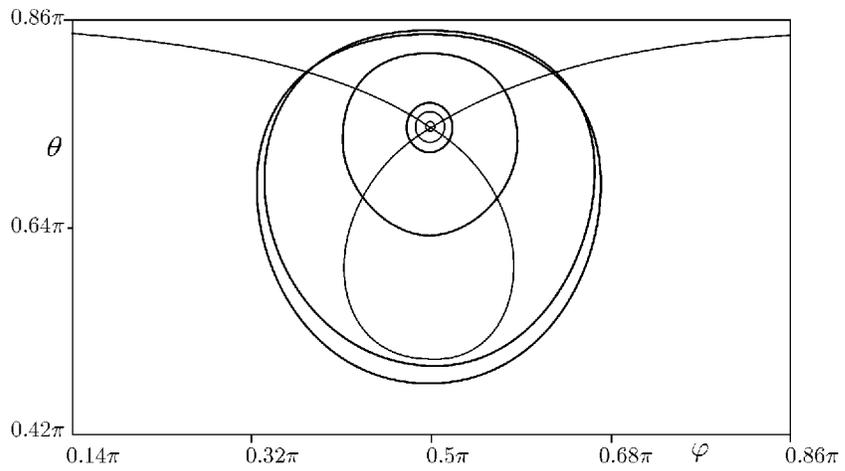}
\caption{Periodic solutions that appear from the saddle point ($\pi/2$, $\theta_2$)
(a part of the sphere is shown in the spherical coordinates). The boundary of the Hill domain
(see Fig. \ref{X1} for $\mathcal{E}=0.5$)  is shown for the critical value of the energy
$E=0$ when other parameters $w=2$, $\gamma=1$. The  values of the energy
corresponding to shown orbits are $E$=0.0009, 0.01, 0.027, 0.355, 1.0, 1.024}
\end{center}
\label{X3}
\end{figure}

\end{document}